\documentstyle[12pt]{article}
\newcommand{\ssu}{SU(2)_L\times SU(2)_R\times U(1)_{B-L}\,}
\newcommand{\matr}{\left( \begin{array}}
\newcommand{\ematr}{\end{array} \right)}
\newcommand{\g}{\gamma}

\newcommand{\dis}{\displaystyle}
\newcommand{\be}{\begin{equation}}
\newcommand{\ee}{\end{equation}}
\newcommand{\ba}{\begin{array}}
\newcommand{\ea}{\end{array}}
\newcommand{\beqa}{\begin{eqnarray}}
\newcommand{\eeqa}{\end{eqnarray}}

\newcommand{\cz}{\cos\zeta}

\newcommand{\dfrac}[2]{{\scriptstyle \frac{#1}{#2}}}

\newcommand{\puoli}{\dfrac{1}{2}}

\newcommand{\lsim}
{{\;\raise0.3ex\hbox{$<$\kern-0.75em\raise-1.1ex\hbox{$\sim$}}\;}}
\newcommand{\gsim}
{{\;\raise0.3ex\hbox{$>$\kern-0.75em\raise-1.1ex\hbox{$\sim$}}\;}}
\newcommand{\ta}[2]{\tau_{#1}^{#2}}

\newcommand{\ula}{\lambda_1\lambda_2\lambda_3}
\newcommand{\G}{\Gamma}

\newcommand{\noi}{\noindent}
\newcommand{\var}{\varepsilon}
\newcommand{\mt}{\mu_1\mu_2\mu_3}
\newcommand{\nt}{\nu_1\nu_2\nu_3}

\begin{document}

\begin{titlepage}

\mbox{}\vspace*{-1cm}\hspace*{8cm}\makebox[7cm][r] {\large
HU-TFT-95-7}

\vfill

\Large

\begin{center} {\bf WW$\gamma$ and WWZ Production in $e^-e^-$
Collisions  in the Left-Right Model}

\bigskip
\normalsize

{{\rm A. Pietil\"a}$^a \:\:$ {\rm and} $\:\:$  {\rm J.
Maalampi}$^b $\\ [15pt]$^a${\it Department of Applied Physics,
University of Turku}\\$^b${\it Department of  Physics, Theory
Division,University of Helsinki}}

{January 1995}

\bigskip

\vfill

\end{center}

\normalsize

\begin{abstract}

We have investigated the three vector boson production in
electron-elect\-ron collisions in the framework of the  left-right
symmetric electroweak model. The process occurs due to  lepton
number violating interactions mediated by Majorana neutrinos and
triplet Higgs scalars. We find that only the reactions with a
heavy gauge  boson pair in the final state, $e^-e^- \rightarrow
W^{-}_2 W^{-}_2\gamma$ and
$e^-e^- \rightarrow W^{-}_2 W^{-}_2Z^0_1$, are phenomenologically
interesting  from the point of view of the Next Linear Collider.
If the mass of $W_2^-$ is 0.5 TeV and the mass of the heavy
Majorana neutrino  of the order of 1 TeV, the cross section of
both reactions is in the range 1 to 10 fb at $\sqrt{s}= 2$ TeV,
depending on the mass of the doubly charged triplet Higgs
$\delta^{--}$, yielding a 1
\% background for the pair production of $W_2^-$. At the
$\delta^{--}$ resonance the cross section can be as large as 1 pb.

\end{abstract}

\end{titlepage}

\section{Introduction}\pagestyle{plain}
\setcounter{page}{2}

According to existing plans  the next linear collider (NLC)
will   operate at the center of mass energy range  of 0.5 -- 2
TeV and  deliver 10
$\rm{fb^{-1}}$ of annual integrated luminosity \cite{Saariselka}.
Such a collider would provide sensitive probes for phenomena
beyond the Standard Model (SM), which are expected to manifest
themselves at the TeV energy scale and whose cross sections often
are in the femtobarn range.

In addition to  the $e^+e^-$ reactions, also the  $e^-e^-$, $e^-
\gamma $ and
$ \gamma \gamma $ collision modes are possible in NLC.  These
will be useful in studying the possible  lepton number
non-conservation. One much studied lepton number violating
reaction is $ e^-e^- \rightarrow W^- W^- $. It was first
discussed by  Rizzo  \cite{Rizzo1} in the context of the "classic"
electroweak left-right  symmetric model (LR-model) of Pati,
Salam, Mohapatra, and Senjanovic \cite{Pati}. Recently this
process has been explored in more detail in the framework of  the
same model by  London and Ng
\cite{London},  the present authors  with their collaborators
\cite{Maalampi1}, \cite{Maalampi2},  and Rizzo \cite{Rizzo2},  as
well as Dicus et al. \cite{Dicus}, and Heusch and Minkowski
\cite{Heusch}.

In the present paper we shall study the next order processes
\beqa  e^-e^- &\rightarrow& W^{-} W^{-}\gamma, \label{WWG} \\
e^-e^- &\rightarrow& W^{-} W^{-}Z^0 \label{WWZ}
\eeqa

\noindent in the framework of the LR-model.
 Here $W$ can be the ordinary charged weak boson (to be denoted
by $W_1$) or the heavy charged boson ($W_2$)  predicted by
LR-model, and $Z$ stands for either the ordinary Z-boson ($Z_1$)
or the heavy neutral boson ($Z_2$) predicted by the model.  The
reactions (\ref{WWG}) and (\ref{WWZ}) could provide us useful
information about several basic interactions of LR-model: gauge
boson - lepton, Higgs - lepton,  Higgs - gauge boson -couplings,
and  gauge boson self-couplings. Some of the couplings involved
do not appear in the pure $W$ pair production. The lowest order
Feynman graphs for the processes $e^-e^-  \rightarrow W^-W^-Z$
are presented in Fig 1. The graphs for the process $e^-e^-
\rightarrow W^-W^-\gamma$ are the same with $Z$ replaced by
$\gamma$ except that there are no photon counterparts for the
graphs 1f, 1g and 1h.

A general feature of the reactions (\ref{WWG}) and (\ref{WWZ})
is  that their cross section is highly suppressed unless the $W$
pair of the final state is the heavy  one, $W_2W_2$. This fact,
true also for the reaction $ e^-e^- \rightarrow W^- W^- $, is
connected to the masses and mixings of the Majorana neutrinos. In
the case of the
 $W_1$ production, the lepton number violation strength is set
mainly by the small mixing between the light and the heavy
neutrino, and if this mixing is extremely small or vanishes, by
the mass of the light neutrino, whereas in the case of the  $W_2$
the mass of the heavy neutrino is relevant. The  processes
$e^-e^- \rightarrow \; W^-_1W^-_1Z_{1}(Z_2,\gamma)$  and $e^-e^-
\rightarrow \; W^-_1W^-_2Z_{1}(Z_2,\gamma)$, although
kinematically favoured, have the cross section clearly below the
femtobarn range (unless there exists a s-channel Higgs resonance
at the relevant energy range)  and are hence phenomenologically
uninteresting.

The reaction $e^-e^- \rightarrow  W_2^- W_2^-\g$ has an
advantage,  compared with the processes $e^-e^- \rightarrow
W_2^-W_2^-Z_{1}(Z_2)$,  of having a lower production threshold
and a larger cross section due to  soft photons.  It also  forms
an important background for the pure $W$ pair production.

The organization of the paper is as follows. In Section 2 we
describe  the basic features of the LR-model. To be
self-contained and to fix our notation we give explicitly all
relevant interaction terms. The cross sections of the reactions
(\ref{WWG}) and (\ref{WWZ}) are derived in Section 3. The
numerical results are presented in Section 4. Section 5 is
devoted to summary and conclusions.

\section{Description of the model}

The matter fields are in  the LR-model are set into left-handed
and  right-handed doublets of the gauge group $\ssu$ \cite{Pati}.
Here we are only  considering leptons, which are accommodated as
follows:
\begin{equation}
 \Psi_L  = \matr{c} \nu_e \\ e^-\ematr_L  = ({\bf 2},{\bf
0},-1),\hspace{10pt}
\Psi_R  = \matr{c} \nu_e \\ e^-\ematr_R  = ({\bf 0},{\bf 2},-1),
\end{equation} and similarly for the muon and tau families.

The breaking of the gauge symmetry, following the chain
$SU(2)_L \times SU(2)_R \times U(1)_{B-L} \rightarrow SU(2)_L
\times U(1)_Y \rightarrow U(1)_{em}$, can be arranged by
introducing  a bidoublet Higgs field

\begin{equation}
\begin{array}{c} {\dis\Phi
=\matr{cc}\phi_1^0&\phi_1^+\\\phi_2^-&\phi_2^0
\ematr = ({\bf 2},{\bf 2},0),}
\end{array}
\end{equation} with the vev given by
\begin{equation}
\begin{array}{c} {\dis<\Phi>
=\frac1{\sqrt{2}}\matr{cc}K_1&0\\0&K_2\ematr,}
\end{array}
\end{equation} and a ''right-handed" triplet field $\Delta_R$

 \begin{equation} {\dis\Delta_R=
\frac{\vec{\delta}_R\cdot\vec{\tau}}{\sqrt{2}} =
\matr{cc}\delta_R^+/\sqrt{2}&\delta_R^{++}\\
\delta_R^0&-\delta_R^+/\sqrt{2}\ematr = ({\bf 1},{\bf 3},2)}
\end{equation} with the vev given by
\begin{equation}
\begin{array}{c} {\dis<\Delta_{R}>
=\frac1{\sqrt{2}}\matr{cc}0&0\\v_{R}&0
\ematr.}
\end{array}
\end{equation}

 The triplet Higgs having both $SU(2)_R$ and $U(1)_{B-L}$ charge
takes care of the first step of the symmetry breaking. Its vacuum
expectation value sets the mass scale of the ''right-handed"
gauge bosons $W_2$ and $Z_2$.
 The experimental lower bounds are $M_{W_2}\gsim 0.65 $ TeV and
$M_{Z_2}\gsim 0.45$ TeV \cite{WZmass}.
 Instead of the triplet one could also use a field transforming
as a doublet under
$SU(2)_R$ and having a non-vanishing $B-L$.  The triplet field
has, however, a virtue which makes it a more  natural choice. It
couples to $|\Delta L| = 2$ lepton currents through  the Yukawa
coupling $ih_R
\Psi_R^TC\tau_2\Delta_R\Psi_R$ giving rise to  Majorana mass
terms for right-handed neutrinos. This leads to the see-saw
mechanism of neutrino masses \cite{seesaw} according to which
there are in each fermion  family two Majorana neutrinos, one
very light ($\nu_1$) and another very heavy ($\nu_2$). The
left-handed and the  right-handed neutrinos are related to these
mass eigenstate Majorana neutrinos
 as follows (assuming no interfamily mixing):
\be
\ba{l}
\nu_L=\frac12(1-\g_5)\left(\nu_1\cos\eta-\nu_2\sin\eta\right),
\nonumber \\
\nu_R=\frac12(1+\g_5)\left(\nu_1\sin\eta+\nu_2\cos\eta\right).
\end{array}
\ee

In addition to the bidoublet and the right-handed triplet Higgses
one often introduces also a ''left-handed" triplet
\begin{equation}
\begin{array}{c} {\dis\Delta_L=
\frac{\vec{\delta}_L\cdot\vec{\tau}}{\sqrt{2}} =
\matr{cc}\delta_L^+/\sqrt{2}&\delta_L^{++}\\
\delta_L^0&-\delta_L^+/\sqrt{2}
\ematr = ({\bf 3},{\bf 1},2),}
\end{array}
\end{equation} and gives it a vev
\begin{equation}
\begin{array}{c} {\dis<\Delta_{L}>
=\frac1{\sqrt{2}}\matr{cc}0&0\\v_{L}&0
\ematr.}
\end{array}
\end{equation}

\noi Phenomenologically it is not, however, necessary, since it
is not needed for the symmetry breaking or the see-saw mechanism.
It will appear in our general expressions for the interactions,
but since its effects in the processes we are considering would
in any case   be small, we will omit it in our later analysis.

There are alltogether seven gauge bosons in the model:
$W_{L,R}^{\pm}$, $W_{L,R}^3$, and $B$. They are related to the
physical massive vector bosons $W_{1,2}$ and
$Z_{1,2}$, and to photon $\gamma\equiv Z_3$ as follows:

\be \matr{c} W_L^\pm \\[5pt] W_R^\pm
\ematr =\matr{cc}
\cos\zeta&-\sin\zeta\\[5pt]\sin\zeta&\cos\zeta\ematr
\matr{c}W_1^\pm\\[5pt]W_2^\pm\ematr
\label{Wmix}\ee and
\begin{equation}
\matr{c}W_L^3\\W_R^3\\B\ematr
=(R_{ij})\matr{c}Z_1\\Z_2\\Z_3=\gamma\ematr\hspace{20pt}(i=L,R,B\:
,\; j=1,2,3).
\label{R}\end{equation}

\noi The general form of the matrix R, as well as the expressions
for the vector boson masses in terms of coupling constants and
the vev's of the Higgs fields, can be found e.g. in \cite{Polak}.
In the limit, where LR-model reproduces the results of the
Standard Model, the mixing matrix
$R$ takes the form

  \be R=
\matr{ccc}
\cos\theta_w& 0&
\sin\theta_w
\\[10pt] -\tan\theta_w\sin\theta_w&
\sqrt{\cos2\theta_w}/\cos\theta_w&
\sin\theta_w
\\[10pt] -\sqrt{\cos2\theta_w}\tan\theta_w& -\tan\theta_w&
\sqrt{\cos2\theta_w}
\ematr,
\label{SMR}
\ee

\noi  where $\theta_w$ is the counterpart of the Weinberg angle.
This corresponds to the case where $M_{W_2},\; M_{Z_2}\to\infty$
and $\zeta\to 0$.

Let us now consider the interactions of the left-right symmetric
model. The following parts of the Lagrangian are involved in the
processes we are interested in:

\noi{\it Neutral current interactions of the electron:}
\beqa {\cal{L}}_{nc}^e &=&
-\puoli\left[g_L\,{\overline{e}}_L\,\gamma^{\mu}\,e_L R_{Ll}+
g_R\,{\overline{e}}_R\,\gamma^{\mu}\,e_R
\,R_{Rl}+g'\,\overline{e}\,
\gamma^{\mu}\,e \,R_{Al}\right] Z_{l\mu}
\nonumber \\  &=& \sum_{l}\overline{e}\,\g^\mu\left(
G^L_{eeZ_l}\,\frac{1-\gamma_5}{2}+
G^R_{eeZ_l}\,\frac{1+\gamma_5}{2}\right)\,e\,Z_{l\mu}
=\sum_{l}\overline{e}\,\G^\mu_{eel}\,e \,Z_{l\mu},
\label{LNCE}
\eeqa where $l=1,2,3.$

\noi{\it Neutral current interactions of neutrinos:}
\beqa {\cal{L}}^{\nu}_{nc} &=&
\puoli\left[g_L\,{\overline{\nu}}_L\,\gamma^{\mu}\,\nu_L \,R_{Ll}+
g_R\,{\overline{\nu}}_R\,\gamma^{\mu}\,\nu_R \,R_{Rl}-g'
\,(\overline{\nu}_L\,
\gamma^{\mu}\,\nu_L+{\overline{\nu}}_R\,\gamma^{\mu}\,
\nu_R)\,R_{Al}\right]Z_{l\mu}  \nonumber \\
&=&\sum_{jj'l}\overline{\nu}_j\,\Gamma_{jj'l}^{\mu}\,\nu_{j'}\,Z_{l\mu},
\label{LNCN} \eeqa where $l=1,2,3$ and $j,j'=1,2$.

\noi{\it Charged current interactions:}
\beqa {\dis{\cal{L}}_{cc}} &=&
{\dis\sum_{jl}{\overline\nu}_j\left(G_{lj}^L\,\g^\mu
\frac{1-\g_5}{2}+G_{lj}^R\,\g^\mu \frac{1+\g_5}{2}\right)
e\,W_{l\mu,-}^\dagger\hspace{5pt}+h. c.}\nonumber \\[10pt]
&=&{\dis\sum_{jl}{\overline\nu}_j\,\Gamma_{lj}^\mu
\,e\,W_{l\mu,-}^\dagger\hspace{10pt}+h. c.}
\label{LNCC}
\eeqa

\noi{\it ZWW interaction:}
\be {\cal{L}}_{ZWW}= i\;\sum_{l,l',l''}e_{l'l''l}
\left(W_{l'\mu\nu}^+W_{l''}^{-\mu}Z_l^\nu-W_{l''\mu\nu}^-W_{l'}^{+\mu}Z_l^\nu
+W_{l'\mu}^+W_{l''\nu}^-Z_l^{\mu\nu}\right),
\label{LZWW}
\ee where we have used the notation
$W_{\mu\nu}=\partial_{\mu}W_{\nu}-\partial_{\nu}W_{\mu}$.

\noi{\it Yukawa interactions:}
 \beqa {\cal{L}}_{Yu} &=& f_\phi \overline{\Psi}_R \Phi
\Psi_L      +g_\phi\overline{\Psi}_R \tilde{\Phi} \Psi_L +ih_L
\Psi_L^TC\tau_2\Delta_L\Psi_L +ih_R \Psi_R^TC\tau_2\Delta_R\Psi_R
+ h.c.
 \nonumber \\   &=& \{{\overline{\nu}}_R e_L [f_\phi\phi_1^+ -
g_\phi\phi_2^+] +{\overline{e}}_R\,\nu_L\,[f_\phi\phi_2^- -
g_\phi\phi_1^-]+h.c\}
\nonumber \\  &+& h_R\{\nu_R^T C\nu_R\delta^0 - e_R^T C
e_R\,\delta^{+\,+} - \sqrt{2}\,\nu_R^T C
e_R\,\delta^{+}+h.c\}+\cdots ,
\label{LYU}
\eeqa where  $\tilde{\Phi}=\sigma^2\Phi^*\sigma^2$.

\noi{\it Kinetic term of the Higgs triplet:}
\beqa {\cal L}_{\Delta}^{kin} &=&
\{({\partial}_{\mu} - i g' B_{\mu} + g \vec{W}_{\mu} \times )
\vec{\delta} \}^{\dagger}
\cdot \{ ({\partial}^{\mu} - i g' B_{\mu} + g \vec{W}^{\mu}
\times )
\vec{\delta} \}
\nonumber \\  &=& g^2\{(\vec{W}_{\mu}\cdot\vec{W}^{\mu})\;
(\vec{\delta}^{\dagger}\cdot\vec{\delta}) - (\vec{W}_{\mu}
\cdot\vec{\delta}^{\dagger})\;(\vec{W}^{\mu}\cdot\vec{\delta})\}+\cdots
;
\label{LKIND}
\eeqa
\noi{\it Kinetic term of the Higgs doublet:}
\be {\cal{L}}_{kin}^{\phi} ={ \rm Tr}\{ (D_{\mu}\Phi)^{\dagger}
(D^{\mu}\Phi)\}
\ee where
\be D_{\mu}\Phi=\partial_\mu\Phi-\frac{i}{2} (g_L\vec{\tau}\cdot
\vec{V}_{L\mu}\Phi - g_R\Phi\vec{\tau}\cdot\vec{V}_{R\mu})
\label{LKINB}
\ee yielding
\be
 -\frac{g_Lg_R}{\sqrt{2}}\cz\{R_{Ll}W^-_ 2 [K_1(\Phi^+_1)^\dagger
- K_2\Phi^-_2 ] + R_{Rl}W_1^- [K_1\Phi_2^- -
K_2(\Phi_1^+)^\dagger]\} Z_l +\cdots \;\;.
\label{LKINF}
\ee

In eqs. (\ref{LKIND}) and (\ref{LKINB}) we have  used the
Cartesian   components of the gauge  bosons
($W^{\pm}_{L,R}=(W^x_{L,R}\mp iW^y_{L,R})/\sqrt{2},$
$W^0_{L,R}=W^z_{L,R}$) and of the triplet Higgses.

The various coupling constants  appearing in the formulas are
defined as follows. The relations between the gauge coupling
constants and the charge $e$ of the positron are given by (the
matrix $R$ is defined in eq. (\ref{R}))
 \be g_L=\frac{e}{R_{L3}},\hspace{10pt}
g_R=\frac{e}{R_{R3}},\hspace{10pt} g'=\frac{e}{R_{B3}}.
\label{gLRB}
\ee The neutral current couplings of the electron are described by
\begin{equation}
\Gamma_{eeZ_l}^\mu=G_{eeZ_l}^L\,\g^\mu \frac{1-\g_5}{2}
+G_{eeZ_l}^R\,\g^\mu \frac{1+\g_5}{2},
\end{equation} where
\begin{equation}
\begin{array}{l}
   G^L_{eeZ_l} = -\frac12 (g_L R_{Ll} + g' R_{Bl}), \\ \\
   G^R_{eeZ_l} = -\frac12 (g_R R_{Rl} + g' R_{Bl}),
\end{array}
\label{GEEZ}
\end{equation} and those of neutrinos by
\be
\begin{array}{c}
\G^{\mu}_{jj'Z_l}= \frac{1}{4}{\g_5 \, \g^{\mu}} \cdot
\left\{
\begin{array}{ll}
 g_L \cos^2 \eta \,R_{Ll} - g_R \sin^2 \eta \,R_{Rl}
 -g' (\cos^2 \eta - \sin^2 \eta ) R_{Bl} ,  &  j=j'=1 \\  [5pt]
-(g_L R_{Ll}+g_R R_{Rl}- 2g' R_{Bl}) \cos\eta
\sin \eta, &   j\neq j' \\ [5pt]
 g_L \sin^2 \eta \,R_{Ll}-g_R \cos^2 \eta \,R_{Rl}
 +g' (\cos^2 \eta - \sin^2 \eta) R_{Bl} ,  &  j=j'=2.
\end{array}
\right.
\\[30pt]
\end{array}
\label{GNNZA}
\ee The charged current vertices are given by
\begin{equation}
\Gamma_{lj}^\mu=G_{lj}^L\,\g^\mu \frac{1-\g_5}{2}
+G_{lj}^R\,\g^\mu \frac{1+\g_5}{2},
\end{equation} where
\be\ba{c} G_{lj}^L=
\frac{1}{\sqrt2}g_L
\matr{cc}
\cos\eta\cos\zeta&\sin\eta\cos\zeta\\
-\cos\eta\sin\zeta&-\sin\eta\sin\zeta
\ematr ,\label{GNWL}\\[10pt] G_{lj}^R=
\frac{1}{\sqrt2}{g_R}
\matr{cc} -\sin\eta\sin\zeta&\cos\eta\sin\zeta\\
-\sin\eta\cos\zeta&\cos\eta\cos\zeta
\ematr, \label{GNWR}
\ea
\ee and finally the $WWZ$ couplings are
\be
\begin{array}{c} - \, e_{ll'l''}=
\left\{
\begin{array}{ll} g_L \,\cos^2\zeta \, R_{Ll''}+ g_R\,\sin^2\zeta
\, R_{Rl''}, &l=l'=1\\[5pt]
\bigl(g_RR_{Rl''}-g_LR_{Ll''}\bigr) \,\sin\zeta\cos\zeta, &l\neq
l' \\[5pt] g_L \,\sin^2\zeta \, R_{Ll''}+ g_R \,\cos^2\zeta
\,R_{Rl''}, &l=l'=2.
\end{array}
\right.
\\[30pt]
\end{array}
\label{ELLL}
\ee

The above equations are given for a general case. We know,
however, that in practice the mixing between charged gauge bosons
and between left- and right-handed neutrinos are small. If we
neglect these mixings alltogether by setting $\eta=\zeta=0$ we
find for the couplings involved in the phenomenologically
interesting reactions $e^-e^-\to W_2^-W_2^-\gamma$ and
$e^-e^-\to W_2^-W_2^-Z$ the following  simple expressions:  \be
\begin{array}{l}
   \Gamma^{\mu}_{22} = \frac{1}{2} G^R_{22}
\gamma^{\mu}(1+\gamma_5)
   =\frac{1}{2\sqrt{2}}g\gamma^{\mu}
   (1+\gamma_5) \equiv G_{enW}\;\gamma^{\mu}(1+\gamma_5),
   \\
   2\;\Gamma^{\mu}_{22l}= \frac{1}{2}
   (g\;R_{Rl}- g'\;R_{Bl})\;\gamma^{\mu}\;\gamma^5 \equiv
   - G_{nnZ}\;\gamma^{\mu}\;\gamma^5,
   \\
   e_{22l} = -\, g_R\;R_{Rl} \equiv \, G_{WWZ}.
\end{array}
\label{kertoimet}
\end{equation} These obey the  relation
\be G_{WWZ} = G^R_{eeZ}+G_{nnZ}.
\ee

A complete analysis of spontaneous symmetry breaking for the
LR-model with one bidoublet, one left-triplet and one
right-triplet Higgs field was presented in ref.
\cite{Deshpande}. It was shown that  the phenomenology  of the
Higgs sector is quite restricted and depends crucially on  three
constants, called $\beta_i$, $i=1,2,3$, appearing in the general
Higgs potential.  If one wants to have neutrinos as Majorana
particles and  preserve the see-saw  mechanism for their masses,
as well as at the same time keep the extra Higgs particles and
gauge bosons light enough to be accesible for the TEV-scale
accelerators,  the couplings
$\beta_i$  should be fine-tuned at least to the order of
$10^{-7}$. To avoid this unnatural situation one could  constrain
$\beta$'s to vanish, e.g. by introducing a suitable extra
symmetry beyond the ordinary LR-model, in which case the mass
scales $v_L$ and $v_R$ are disconnected and there remains  a
remnant see-saw relation, which is most naturally satisfied by
the  condition $v_L=0$ \cite{Deshpande}.

In the case that the vev $v_L$ of the left-handed triplet
$\Delta_L$ vanishes (or if $\Delta_L$ does not exist at all), the
linear  combination
$K_1\,\phi_2^- \,-\,K_2 \,\phi_1^-$  is the Goldstone field
corresponding to the longitudinal component of the light weak
boson $W_1^-$. The Goldstone field corresponding the longitudinal
component of the heavy gauge boson $W_2^-$ is in turn the
superposition
$\delta^-_R- (K_1^2-K_2^2)\phi^-_1/\sqrt{2}K_1v_R$. Hence when
the left-handed triplet is neglected the only  physical singly
charged  Higgs field $h^-$ is

\be h^-=
\frac{1}{\sqrt{1+K^2_1/2v^2_R(\frac{K_1^2-K_2^2}{K_1^2+K_2^2})^2}}(\phi^-_1
+
\frac{K_1}{\sqrt{2}\; v_R}
\frac{K_1^2-K_2^2}{K_1^2+K_2^2}\delta^-_R). \ee

\noi The doubly charged Higgs $\delta^{--}_R$ is, of course, a
physical field as it does not mix with any other field and is not
eaten during the spontaneous symmetry breaking. Both $h^-$ and
$\delta^{--}$ have lepton number violating interactions and they
contribute in the processes we are considering.

The mass of the electron $m_e$ and  the Dirac mass  of the
neutrino $m_D$  are given by the relations
\be
\begin{array}{l} m_e = (f_\phi K_2+g_\phi K_1)/{\sqrt{2}},
\nonumber \\  m_D = (f_\phi K_1+g_\phi K_2)/{\sqrt{2}}.
\end{array}
\end{equation} Both of them are very small compared with the
considered particle energies and can be safetely neglected in our
calculations.  According to the see-saw mechanism
$m_D$ is related to the masses $m_{\nu_1}$ and
$m_{\nu_2}$ and to the mixing angle $\eta$  of the Majorana
neutrinos $\nu_1$ and $\nu_2$, assuming no interfamily mixing, in
the following way:

\be m_D = \puoli (m_{\nu_2}-m_{\nu_1}) \sin 2\eta,
\ee where the mixing angle $\eta$ is given by

\be
\tan{\eta}=\sqrt{\frac{m_{\nu_1}}{m_{\nu_2}}}.
\ee The heavy neutrino mass is most naturally in the TeV range
and hence the mixing angle is of the order of $10^{-6}$. This
indicates that $m_D$ would be in the MeV scale and its omission
is justified.

The Feynman rules for the vertices needed for the calculation of
the lowest  order amplitudes of the reactions (\ref{WWG}) and
(\ref{WWZ}) are collected in  Fig. 2.

\section{Amplitudes }

In this section we shall present the interaction amplitudes of
the processes (\ref{WWG}) and  (\ref{WWZ}). The Feynman diagrams
corresponding the latter reaction are given in Fig. 1. The
graphs for the former reaction are the same with the $Z$-boson
replaced by the photon, except that there are no counterparts of
the diagrams 1f, 1g and 1h in the   photon case.

As was mentioned in Introduction, the cross sections are in a
phenomenologically interesting range only if both of the final
state W-bosons are the heavy ones, $W_2$. It is easy to convince
oneself about this by considering, for example, the diagram 1b.
If the final state W-boson denoted by $W(2)$ is the light boson
$W_1$, which couples  mainly  in V--A currents,   the lower $e
\nu W$ vertex is suppressed in the case of the heavy neutrino
$\nu_2$, whose interactions are mainly of V+A -type. The vertex
favours the light neutrino $\nu_1$, which would mean that in the
upper $e \nu W$ vertex the production of a virtual  $W_1$ is
favoured over that of a virtual $W_2$. The final state $W$-boson
denoted by $W(1)$ would then in the favourable case be $W_1$.
Nevertheless, this situation of unsuppressed couplings would
yield a neglicible cross section because the amplitude is
proportional to the mass of the light neutrino. This is because
of the chirality matching  in the  $e \nu W$ vertices.
 Otherwise  two (or one) light $W$'s in the final state only
follows  from the mixing of the neutrinos or/and the mixing of
the $W$ bosons; every light $W$ of the final state then yields
either the factor $ \sin\eta $ or
$ \sin\zeta $ in the amplitude. As mentioned earlier, according
to the see-saw mechanism the neutrino mixing angle obeys  $
\sin\eta\lsim 10^{-5}
$, and the experimental data gives the constraint for the mixing
angle of charged weak bosons of $ \sin\zeta\lsim 10^{-3} $
\cite{zeta}. Thus in any case the cross sections of the reactions
with one or two $W_1$'s in the final state are very small.

In the following we will thus consider only the reactions with
two $W_2$'s in the final state. The amplitudes for these are
insensitive to the  mixing angles $\eta$ and
$\zeta$. We shall   let them to vanish, which  simplifies our
expressions considerably.  All terms  left in the amplitude are
then proportional to the  heavy neutrino mass $m_{2}$.

For Majorana neutrino fields appearing in the amplitudes we apply
the Feynman rules given in ref. \cite{Gluza}.  From the  graph 1a
and the corresponding crossed graphs we obtain the amplitudes
\be \ba{l} T_1 = c_R\;G_{eeZ}^{R}\cdot\ta{2}{\mu_1\mu_2\mu_3}
(p_1-k_3) / [t_{a3}(u_2-m_{\nu_2}^2)],
\nonumber \\ \nonumber \\ T_2 =
c_R\;G_{eeZ}^{R}\cdot\ta{1}{\mu_3\mu_1\mu_2} (p_2-k_3) /
[t_{b3}(u_1-m_{\nu_2}^2)],
\nonumber \\ \nonumber \\ T_3 =
c_R\;G_{eeZ}^{R}\cdot\ta{2}{\mu_2\mu_1\mu_3} (p_1-k_3) /
[t_{a3}(t_2-m_{\nu_2}^2)],
\nonumber \\ \nonumber \\ T_4 =
c_R\;G_{eeZ}^{R}\cdot\ta{1}{\mu_3\mu_2\mu_1} (p_2-k_3) /
[t_{b3}(t_1-m_{\nu_2}^2)].
\ea
\label{1}
\ee We have used here the notations
\be\ba{l} c_R = - \, 2\, m_{\nu_2}\;G_{enW}^2 \;(1+\g_5) =
		 - \dfrac{1}{4} \,m_{\nu_2}\,g_R^2 \,(1+\g_5),\nonumber\\
   \tau_1^{\lambda_1 \lambda_2 \lambda_3}(q_1)=
   \g^{\lambda_1}\not q_1\g^{\lambda_2}\g^{\lambda_3},
   \nonumber \\
   \tau_2^{\lambda_1 \lambda_2 \lambda_3}(q_2)=
   \g^{\lambda_1}\g^{\lambda_2}\not q_2\g^{\lambda_3}.
   \end{array} \label{TFORM3}
   \end{equation}

The diagrams 1b and 1d and their crossed counterparts, all with a
virtual
$W$-boson decaying into a $WZ$ pair, lead to the amplitudes

 \be
\ba{l} T_5 = -\, { c_R\; G_{WWZ}}
 D^{W_2}_{\mu\nu}(k_1+k_3)F^{\mu_1\nu\mu_3}
 (k_1,k_1-k_3,k_3)\g^{\mu}\g^{\mu_2}/({u_1-m_{\nu_2}^2} ),
\nonumber \\ \nonumber \\ T_6 = -\, { c_R\; G_{WWZ}}
 D^{W_2}_{\mu\nu}(k_2+k_3)F^{\mu_1\nu\mu_3}
 (k_2,k_2-k_3,k_3)\g^{\mu}\g^{\mu_1}/({t_1-m_{\nu_2}^2}),
\nonumber \\ \nonumber \\ T_7 = -\,{ c_R\; G_{WWZ}}
 D^{W_2}_{\mu\nu}(k_1+k_3)F^{\mu_1\nu\mu_3}
 (k_1,k_1-k_3,k_3)\g^{\mu_2}\g^{\mu}/({t_2-m_{\nu_2}^2}),
\nonumber \\ \nonumber \\ T_8 = -\,{ c_R\; G_{WWZ}}
 D^{W_2}_{\mu\nu}(k_2+k_3)F^{\mu_2\nu\mu_3}
 (k_2,k_2-k_3,k_3)\g^{\mu_1}\g^{\mu}/({u_2-m^2_2}),
\nonumber \\ \nonumber \\ T_9 = -\, {4c_R \; G_{WWZ}}
D^{W_2}_{\mu\nu}(k_1+k_3)F^{\mu_1\nu\mu_3}
 (k_1,k_1-k_3,k_3)g^{\mu\mu_2}/({s-M_{\delta}^2}),
\nonumber \\ \nonumber \\ T_{10} = -\, {4c_R \; G_{WWZ}}
D^{W_2}_{\mu\nu}(k_2+k_3)F^{\mu_2\nu\mu_3}
 (k_2,k_2-k_3,k_3)g^{\mu\mu_1}/({s-M_{\delta}^2}).
\ea
\ee
\label{2a} Here we have defined
\be F^{\ula}(q_1,q_2,q_3) =
2q_3^{\lambda_1}g^{\lambda_2\lambda_3}+
q_2^{\lambda_2}g^{\lambda_1\lambda_3}-2q_1^{\lambda_3}g^{\lambda_1\lambda_2}
\label{FQQQ}
\ee and
\be D^{W}_{\mu\nu}(k) = -\left(g_{\mu\nu}-\frac{k_{\mu}k_{\nu}}
{M^{2}_{W}}\right) / \left(k^2-M^{2}_{W}\right).
\ee {}From the graphs 1c and 1e and the corresponding crossed
graphs, which correspond to production of a
$Z\delta^{--}$ pair followed by a $\delta^{--}$ decay into a $WW$
pair, one obtains the amplitudes

\be
\begin{array}{ll} T_{11} &= 4c_R\;G^R_{eeZ}\;{g^{\mu_1\mu_2}}
\; {\gamma^{\mu_3}(\not p_2-\not
k_3)}/[{t_{b3}}({s_3-M_{\delta}^2})],
 \\  \\ T_{12} &= 4c_R\;G^R_{eeZ}\;{g^{\mu_1\mu_2}}\; {(\not
p_1-\not k_3)\gamma^{\mu_3}}/[{t_{a3}}({s_3-M_{\delta}^2})],
 \\ \\ T_{13} &= 16c_R\;G^R_{eeZ}\;{g^{\mu_1\mu_2}}\;
{(p_1+p_2)^{\mu_3}}/[({s_3-M_{\delta}^2})({s-M_{\delta}^2})].
\end{array}
\label{3}
\ee

In the photon case one obtains similar amplitudes with $G_{WWZ}$
and
$G_{eeZ}$ replaced with $-e$.
 The graph 1f and its crossed graph, where  Z is produced in  the
virtual neutrino line, leads to the amplitudes
\be
\ba{l} T_{14} = c_R\;G_{nnZ}
[\ta{1}{\mu_2\mu_3\mu_1}(p_2-k_2)+\ta{2}{\mu_2\mu_3\mu_1}
(p_1-k_1)] / [(t_1-m_{\nu_2}^2)(t_2-m_{\nu_2}^2)],
\nonumber \\ \nonumber \\ T_{15} = c_R\;G_{nnZ}
[\ta{1}{\mu_1\mu_3\mu_2}(p_2-k_1)+\ta{2}{\mu_1\mu_3\mu_2}
(p_1-k_2)] / [(u_1-m_{\nu_2}^2)(u_2-m_{\nu_2}^2)],
\ea
\label{4}
\ee

\noi and the graphs 1g and 1h and their crossed graphs,
involving a virtual Higgs  $h^-$, lead to
\be
\ba{l} T_{16} = 2c_R\;G_{hZ}\;{g^{\mu_2\mu_3}}\;
{\gamma^{\mu_1}(\not p_2-\not
k_1)}/[({s_2-M_h^2})({u_2-m_{\nu_2}^2})],
\nonumber \\ \nonumber \\ T_{17} =
2c_R\;G_{hZ}\;{g^{\mu_2\mu_3}}\; {(\not p_1-\not
k_1)\gamma^{\mu_1}}/[({s_2-M_h^2})({t_1-m_{\nu_2}^2})],
\nonumber \\ \nonumber \\ T_{18} =
8c_R\;G_{hZ}\;{g^{\mu_2\mu_3}}\;
{(p_1+p_2)^{\mu_1}}/[({s_2-M_h^2})({s-M_{\delta}^2})],
\nonumber \\ \nonumber \\ T_{19} =
2c_R\;G_{hZ}\;{g^{\mu_1\mu_3}}\; {\gamma^{\mu_2}(\not p_2-\not
k_2)}/[({s_1-M_h^2})({t_2-m_{\nu_2}^2})],
\nonumber \\ \nonumber \\ T_{20} =
2c_R\;G_{hZ}\;{g^{\mu_1\mu_3}}\; {(\not p_1-\not
k_2)\gamma^{\mu_2}}/[({s_1-M_h^2})({u_1-m_{\nu_2}^2})],
\nonumber \\ \nonumber \\ T_{21} =
8c_R\;G_{hZ}\;{g^{\mu_1\mu_3}}\;
{(p_1+p_2)^{\mu_2}}/[({s_1-M_h^2})({s-M_{\delta}^2})].
\ea
\label{5}
\ee In these equations we have used the notation
\be G_{hZ} = \frac{(K_1/\sqrt{2}v_R)^2}{1+(K_1/\sqrt{2}v_R)^2} \;
		(g_L R_{Ll} + g_R R_{Rl} - 2g' R_{Bl}).
\label{GHZ}
\ee

Let us note incidentally that the amplitudes $T_{16}, ...,T_{21}$
are of the same form as the terms arising from the longitudinal
part of the W-propagator in the amplitudes $T_{7}, ..., T_{10}$,
because, e.g.,
$D^{W_2}_{\mu\nu}(k_1+k_3)F^{\mu_1\nu\mu_3}
 (k_1,k_1-k_3,k_3)\g^{\mu_2}\g^{\mu}$ is equivalent to
$F^{\mu_1\nu\mu_3}
 (k_1,(p_2-k_2)(M_3/M_2)^2 - 2k_3,k_3)\g^{\mu_2}\g^{\mu}$ when
multiplied by the relevant polarization vectors. In other words,
they just modify the expressions $(M_3/M_2)^2/(s_1-M_2^2)$ and
$(M_3/M_2)^2/(s_2-M_2^2)$ appearing in $T_5,..., T_{10}$.

  In eqs. (\ref{2a}), (\ref{3}), and (\ref{5}) the coupling $h_R$
has been  eliminated by using the relation
\be  m_{2} \approx 2 \,h_R \,\langle \delta_R^0 \rangle
=\sqrt{2}\,h_R\,v_R .
\ee

The kinematical variables $s_1$, $s_2$, $t_1$, $t_2$ appearing in
the above formulas  are defined by

\be
\ba{l} s = (p_1+p_2)^2,
\\  s_1 = (k_1+k_3)^2,
\\  s_2 = (k_2+k_3)^2,
\\  t_1 = (p_1-k_1)^2,
\\  t_2 = (p_2-k_2)^2.
\ea
\label{ST}
\ee They are the same variables as given by
 Byckling and Kajantie in \cite{Byckling} for a general $2\to 3$
reaction, except that we have renamed their momenta
 $p_a$, $p_b$, $p_1$, $p_2$, $p_3$  as $p_1$, $p_2$, $k_1$,
$k_3$, $k_2$,  respectively. We have also introduced the
following auxiliary variables:
\be
\ba{lclcl} t_{a3}&=&{(p_1-k_3)}^2 &=& t_2-t_1-s_1+M^2_1+M^2_3,
\\ t_{b3}&=&{(p_2-k_3)}^2 &=& t_1-t_2-s_2+M^2_2+M^2_3,
\\ s_3&=&{(k_1+k_2)}^2 &=& s-s_1-s_2+M^2_1+M^2_2+M^2_3,
\\ u_1&=&{(p_1-k_2)}^2 &=& s_1-t_2-s+M^2_2,
\\ u_2&=&{(p_2-k_1)}^2 &=& s_2-t_1-s+M^2_1,
\ea
\label{SIVU}
\ee where $M_1$, $M_2$, and $M_3$ denote generally the masses of
the gauge bosons $W_{l_1}$,
$W_{l_2}$, and $Z_{l_3}$, respectively.

The complete scattering amplitude is of the form
\be
   M
=i^5\overline{v}(e_2)\;T^{\mt}u(e_1)\;\var_{\mu_1}^*(W_{l_1})\;
   \var_{\mu_2}^*(W_{l_2})\;\var_{\mu_3}^*(Z_{l_3}),
\label{M}
\ee where $T$ is the sum of the amplitudes $T_i$ and $\epsilon$'s
are the polarization vectors of the weak bosons. The unpolarized
total cross section  is then given by the formula

\begin{equation}
  \sigma = {1 \over (2 \pi )^5} {{{\int \prod_{i=1}^3 {{d^3 k_i}
\over
  {2 E_i}} \delta^4 (p_1+ p_2- \sum k_i)\langle  |M|^2\rangle}}
  \over {2 s}}
\label{ALA}
\end{equation} with
\be{ \langle |M|^2 \rangle =\frac{1}{4}\sum_{spins}|M|^2}.
\ee In the case of the reaction $e^-e^-\to W^-W^-Z$ one has

\begin{eqnarray}
\langle|M|^2\rangle &=&-\frac{1}{4}(g^{\mu_1\nu_1}-
   \frac{k_1^{\mu_1}k_1^{\nu_1}}{M_1^2})\;
   (g^{\mu_2\nu_2}-\frac{k_2^{\mu_2}k_2^{\nu_2}}{M_2^2})\;
   (g^{\mu_3\nu_3}-\frac{k_3^{\mu_3}k_3^{\nu_3}}{M_3^2})
   \nonumber \\
   & & \times {\rm Tr}\;\{T_{\mt}\not p_1\g^0
{(T_{\nt})}^{\dagger}\g^0\not p_2\}, \label{MNELIO}
\end{eqnarray}

\noi  in the case of the reaction $e^-e^-\to W^-W^-\gamma$ one
replaces
$-(g^{\mu_3\nu_3}-{k_3^{\mu_3}k_3^{\nu_3}}/{M_3^2})$ with
$-g^{\mu_3\nu_3}$.
 If there are two identical bosons in the final state,   the
expression (\ref{ALA}) should be multiplied by one half.

A typical feature of the gauge theories is a delicate
cancellations among the different partial amplitudes which
guarantee a good high-energy behaviour of the total cross
section. This offers a good check of the calculation. Another
cross check is gauge invariance of the total amplitude, according
to which the amplitude $M$ in the case of
 a photon  should vanish when one replaces the polarization
vector   $\var_{\mu_3}^*(\g)$ with the photon momentum
$k_3^{\mu_3}$ and performs the contraction.  Our results pass this
check when we take  into account that  we have neglected the
electron mass.

We have given our formulas in a form where
 the width of $\Delta^{--}$ is neglected. Including the width is,
of course, straightforward and would  not remarkably complicate
our computations.  The width  can be evaluated by using the
formulas
\beqa
\Gamma_{\Delta^{--} \rightarrow l^-l^-}&=& {g_R^2
M_{\Delta}}(M_{\nu e}^2  + M_{\nu \mu}^2 + M_{\nu \tau}^2)/({32
\pi }{M_{W_2}^2}),
\nonumber\\
\Gamma_{\Delta^{--} \rightarrow W_2^-W_2^-}&=& \frac{g_R^2
M_{\Delta}}{4\pi}
\; \sqrt{1 - 4 ({M_{W_2}}/{M_{\Delta}})^2} \; [1 - ({ M_{W_1}}/{
M_{W_2} })^2] \nonumber\\ &\phantom{xx} &\times [3 ({ M_{W_2}}/{
M_{\Delta}})^2 +  ({M_{\Delta}}/{2M_{W_2}})^2 -1],
\\
\Gamma_{\Delta^{--} \rightarrow W_2^- h^-} &=&
\frac{g_R^2 M_{\Delta}}{4\pi} \; ({M_{W_1} M_{\Delta}}/{
M_{W_2}^2})^2
\;
\lambda^{3/2}[1,({M_{W_2}}/{M_{\Delta}})^2,({M_h}/{M_{\Delta}})^2]\;.\nonumber
\label{wid}
\eeqa The two first formulas  can be infered from \cite{Gunion},
which gives the corresponding results for the left-handed
triplet.  The third formula can be deduced from their result by
expressing the triplet field
$\Delta^-$  in terms of the mass eigenstate
$h^-$ and the appropriate Goldstone field. We have assumed $h$ so
heavy  that the contribution of the third channel to the total
width can be neglected \cite{Deshpande}.

\section{Numerical results}

Let us now describe our computations. The trace manipulations in
the squared matrix element (\ref{MNELIO}) were  carried out by
using the symbolic manipulation program REDUCE and they were
checked with MATHEMATICA. The phase space integration in
(\ref{ALA}) was performed by using the formula
\cite{Byckling}

\be {\dis\int_{\rm phase\;space}<|M|^2> \;=
\frac{\pi}{16s}\;\int<|M|^2>
\frac{ds_2\,ds_1\,dt_1\,dt_2}{\sqrt{-\Delta_4}}},
\ee where
\be 16\cdot\Delta_4 =
\left|
\begin{array}{cccc} 0 & s & M_1^2-t_1 & t_2+s-s_1
\\ s & 0 & t_1+s-s_2 & M_2^2-t_2
\\ M_1^2-t_1 & t_1+s-s_2 & 2M_1^2 & s-s_1-s_2+M_3^2
\\ t_2+s-s_1 & M_2^2-t_2 & s-s_1-s_2+M_3^2 & 2M_2^2
\end{array} \right|.
\ee The inner integration in $t_1t_2$-plane is over an ellipse
defined by the condition $\Delta_4 = 0$. (We performed this
integration also analytically for a check.)   The 4-fold
integrals were performed numerically by using  the Monte Carlo
program VEGAS \cite{Lepage}.

The symmetry relations
\be
\ba{l}
\sigma_{ij} = \sigma^*_{ji}\; ,
\nonumber  \\
\sigma_{ij} = \sigma_{i_cj_c}
\ea
\label{symrel}
\ee turned out to be  of help in  checking  the code. Here the
$\sigma_{ij}$ is the cross section contribution coming from the
interference term of the amplitudes $T_i$ and $T_j$, and $i_c$
refers to the   amplitude obtained from the amplitude of the
index $i$ by charge  conjugation:
\be T_{i_c}(p_2,p_1) = C\;T^T_i(p_1,p_2)\;C^{-1}.
\ee We found out that the symmetry relations (\ref{symrel}) are
in general numerically satisfied to several digits although the
values of $\sigma$'s themselves vary somewhat with the Monte
Carlo parameters used.

The input of our computation consists of the particle masses, the
gauge coupling constants, and various mixing angles. Assuming
that there is no mixing between different neutrino flavours, only
the neutrino mixing angle $\eta$ between the left-handed and the
right-handed electron neutrino enters our calculations.  The
mixing between $W_L$ and $W_R$  is described with the angle
$\zeta$ as presented in eq. (\ref{Wmix}) . The mixing between the
neutral gauge bosons is described in terms of three mixing
angles, out of which only two are independent in the case that
$g_L=g_R$ as we shall assume. The mixing matrix, which we have
denoted $R$ in the foregoing section, can then be determined  by
using the experimental data for the two neutral current
parameters, e.g. the vector and axial vector $Z_1$ couplings of
quarks and leptons \cite{Maalampi1}. In our calculations we have
used  three different forms for it: the SM limit of R given by
(\ref{SMR}) where the Weinberg-Salam angle $\theta_{WS}$ has been
estimated from the experimental value of $M_{W_1}/M_{Z_1}$, the
numerical form obtained as in \cite{Maalampi1}, and the numerical
form calculated by using the matrix given in \cite{Rizzo5} and
fixing the $M_{W_1}$ and $M_{Z_1}$ to their experimental values.

For the mass of the heavy W-boson we use the value 0.5 TeV
(sometime also the value 0.7 TeV for comparison), which is close
to its experimental lower bound.  For the mass of $\delta^{--}$
Higgs we have used two values, one below the reaction threshold
($M_{\delta}=0.8 $ TeV), another one well above the threshold
($M_{\delta}=10$ TeV).

 The cross section is proportional to the square of the mass of
the heavy neutrino for which we used the value
$m_{2}=1$ TeV. This is a natural choice in the sense that the mass
originates in the breaking of the $SU(2)_R$ symmetry, which we
have assumed to occur at the TeV-scale. Letting the Yukawa
coupling $h_R$ differ from the value $O(1)$, also much lighter
neutrino mass were possible, but then also the cross section were
much smaller.  At the energies considerably above the neutrino
mass the cross section roughly scales with $m_{2}^2$.

As mentioned, there are delicated cancellations among the
amplitudes contributing to the processes we are considering. We
demonstrate this  in Fig 3 where we have plotted separately cross
sections for  the reaction $e^-e^-\to W_2W_2Z_1$ as a function of
the collision energy
$E_{cm}=\sqrt{s}$ corresponding to the subsets of the amplitudes
given in  eqs. (\ref{1}), (\ref{2a}), and (\ref{3}) and their
interferences.  The contribution from the amplitudes (\ref{4}) is
negligibly small, and  it was, like the suppressed amplitudes
(\ref{5}), omitted in the Fig 3.  Here, and in what follows if
not otherwise stated,  we have taken the mass of $W_2$ equal to
0.5 TeV and the mass of the heavy neutrino equal to 1 TeV. The
triplet Higgs mass is in this figure equal to 10 TeV. The total
cross section, invisible in the scale of the figure, is some
thousands of the partial contributions plotted in the figure. At
$E_{\rm cm}$ = 2 TeV we obtain $\sigma_{\rm  tot}=2.4$ fb. The
$\sigma_{\rm  tot}$ increases with the collision energy up to
$E_{\rm cm}=M_{\delta}$, after which it starts to decrease
because of a destructive interference of the amplitudes involving
virtual triplet Higgses.

In Fig. 4 we present $\sigma_{\rm tot}$ for $e^-e^-\to W_2W_2Z_1$
as a function of $E_{\rm cm}$ for the case of a light
$\delta^{--}$, $M_{\delta}= 0.8$ TeV and  for the case of a heavy
$\delta^{--}$, $M_{\delta}= 10$ TeV by using the SM limit of R
(solid line) and in the latter case also by using  the R given in
\cite{Maalampi1} (dashed line) and the R which is  the most
consistent with the model
\cite{Rizzo5} (dashdot line).  In the case of $M_{\delta}= 0.8$
TeV the cross section shows a  maximum of 4.5 fb at around  2
TeV. With the anticipated luminocity $10$ fb$^{-1}$ the peak value
would correspond to 45 events per year.  The results obtained by
using the three different choices of the R matrix are very similar
at low energy region (all presented by the divide line in the
figure),
 but they start to differ at high energies. The cross section
corresponding to the SM limit of R and to R  of \cite{Rizzo5} has
the expected good  high-energy behaviour, while in the case of
our third R the cancellations among various amplitudes are less
complete.

In Fig. 5 we present the total cross section for the reaction
$e^-e^-\to W_2W_2\gamma$ (here $M_{\delta}= 0.8$ TeV) . A new
feature compared with the $WWZ$  case is  the singularity caused
by soft-photons.  We handle  the singularity following the
standard method of  evading it by applying cuts  for the minimum
energy $E_{\gamma}$ and the minimum scattering angle
$\theta_{\gamma}$ of the photon,  which is a natural procedure
from the experimental point of view. In Fig. 5 we have used the
cuts $\cos\theta_{\gamma}\leq 0.8$ and $E_{\gamma}/E_{\rm cm}\geq
0.01$.

Also here we have demonstrated the effect of cancellations. We
have separated the amplitudes  mediated by the neutrinos  from
those mediated by $\delta^{--}$ Higgses. Both contributions are
separately physical in the sense that their amplitudes are gauge
invariant by themselves. However, they both increase with the
energy contradicting eventually with unitarity.  Their
destructive interference takes care of  the good high  energy
behaviour of the total cross section.

The sensitivity of the total cross section in the applied cuts  is
illustrated in Fig. 6. We present the total cross section for
three sets of the photon  energy and scattering angle cuts. Near
the threshold the cross the effect of changing the cuts is quite
dramatic. The peak value of the total cross section is of the
order of 10 fb, that is,  about two times the cross section in
the case of  the $W_2W_2Z_1$ final state.

Fig. 7 presents the cross section as a function of energy for a
slightly heavier $W_2$, $M_{W_2}=0.7$ TeV. One  notes a large
decrease of  the maximum value of the cross section,  to some
20th part of its value in the case of $M_{W_2}=0.5$ TeV . With
the increasing energy the  cross section approaches to the value
expected according to the scaling with the
$W_2$ mass at high energies.

We note that the finite width of $\delta$ does not essentially
change the results presented above, because the dominant decay
channel of $\delta$ is into two leptons  and this is still
relatively small for the $\delta^{--}$ mass of 0.8 TeV.

The Fig. 8 gives the cross section for $e^-e^-\to W_2W_2\gamma$
in the case of  $M_{\delta}= 10$ TeV. It should be compared with
Fig. 5 where $M_{\delta}= 0.8$ TeV. We have plotted again
separately the contributions from neutrino mediated and the
triplet Higgs mediated amplitudes. In contrast with the  case of
Fig. 5, these contributions interfere now constuctively.

In the case of heavy triplet Higgs, the zero width approximation
applied above is not any more reliable. In fact, for  a mass so
high as 10 TeV the decay $ \delta^{--} \rightarrow W_2^- \;W_2^-
$ alone gives  rise to a width which is greater than the mass of
the Higgs, making the particle interpretation of the Higgs
questionable.  If we nevertheless trust on our perturbational
calculations the total cross section at the experimentally
relevant energy of 2 TeV and below is in a good approximation
given by the pure neutrino contribution, i.e., by the  curve
$\sigma_{n \times n}$   in Fig. 8.

In Fig. 9 we present the total cross section for two sets of the
photon cuts assuming a $\delta$-pole at 2 TeV with a width of
$\Gamma=0.14 M_{\delta}$. The width has been estimated by using
eq. (\ref{wid}) assuming the electron-, muon- and tau-type heavy
neutrinos and $W_2$ to have mass equal to 1 TeV and the Higgs
$h^-$ so heavy that it can be neglected.

In the calculation we made some simplifying assumptions. Firstly,
we have neglected possible family mixing among neutrinos. Such a
mixing might increase the production rates of the final states
consisting of a light $W$ pair or a light and heavy $W$ pair, as
discussed in \cite{Heusch}. Secondly, we have assumed that the
gauge couplings associated with the subgroups $SU(2)_L$ and
$SU(2)_R$ are equal,  $g_L=g_R$. Combined with the assumption of
identical  Cabibbo-Kobayashi-Maskawa matrices for the left- and
right-handed quarks, this is known to  lead to the upper bound
of  1.4 - 1.6 TeV for $M_{W_2}$ \cite{zeta}. This  would push the
production  of the  $W_2$ pair out of the reach of NLC. The
possible deviations  of the ratio $K=g_R/g_L$ from the value one
has been discussed in \cite{Kayser} and \cite{Rizzo4}, where the
allowed region 0.55 to  1.5 was found. The GUT extensions of
LR-model favours the value 1 for the upper limit REF. The cross
sections we have calculated have an overall factor $g_R^4$, but
this hardly leads to the scaling by this factor, because changing
of $K$  change   the couplings $G_{WWZ}$, $G_{eeZ}^R$, and
$G_{nnZ}$, and also  indirectly the  matrix $R$. We do not expect
the value of $K$ essentially change our numerical results at the
energy range of NLC, but at higher energies the effect may be
substantial.

\section{Summary}
 The left-right symmetric electroweak model predicts the
existence of  heavy gauge bosons $W_2^{\pm}$ and $Z_2^0$, as well
as lepton number violating interactions mediated by a heavy
Majorana neutrino and a right-handed triplet scalars. We have
calculated the total cross section for the lepton number
violating processes $e^-e^-\to W_2W_2\gamma$ and
$e^-e^-\to W_2W_2Z_1$, which could be experimentally measured at
NLC provided that $W_2$ is not much heavier than the present
lower limit  of its mass. The reactions are  background
processes for the heavy vector boson pair production $e^-e^-\to
W_2W_2$ investigated earlier
\cite{London},\cite{Maalampi1},\cite{Dicus},
\cite{Heusch}. The cross sections of the reactions where one or
both of the charged bosons in the final state is the ordinary
weak boson
$W_1$ is found to be too small to have any phenomenological
interest as far as NLC in concerned.

If the mass of the heavy electron neutrino is of order of 1 TeV
and  the $W_2$ mass near to 0.5 TeV, we find the total cross
section of
$e^-e^{-}\to W_2^-W_2^- Z_1 $ to be in the collision energy
region of a few TeV is in the range of 1 to 10 fb depending on
the mass of the  double charged triplet Higgs $\delta^{--}$.  The
cross section of
$e^-e^{-}\to  W_2^-W_2^-\g $ under the same assumptions and with
reasonable cuts on the photon energy and scattering angle is of
the same order. Both cross sections increase with increasing
heavy neutrino mass. At the
$\delta^{--}$ pole the cross sections can reach the value 1 pb.
{}From these results one can conclude that $e^-e^{-}\to
W_2^-W_2^-\g $ yields    of the order of 1 \%  background for the
pair production $e^-e^-\to W_2W_2$.

\vspace{1cm}
\Large Acknowledgements
\normalsize
\vspace{1cm}

The authors thank Mr. P. Iljin and Mr. Juha Vuori for their
assistance  in numerical work.  One of the authors (A.P.)
expresses his gratitude to Reseach Institute for High Energy
Physics, Helsinki, where this work was completed, for
hospitality. The work has been financially supported by  Turun
yliopistos\"a\"ati\"o and the Academy of Finland.

\newpage
\noi{\bf Figure captions}

\noi{\bf Figure 1.} The lowest-order Feynman graphs for the
process $e^-e^-\rightarrow W^-W^-Z$.

\noi{\bf Figure 2.} The Feynman rules for vertices appearing in
the amplitudes. Here
$R=(1+\g_5)/2$ and $L=(1-\g_5)/2$, and the other quantities are
defined in the text.

\noi{\bf Figure 3.} The  contributions of various subsets of
amplitudes and their interferences to the cross section of the
reaction $e^-e^-\to W_2^-W_2^-Z_1$ as functions of the CM-energy
$E=\sqrt{s}$ for the masses
$M_{W_2}=0.5$ TeV, $M_{\delta}=10$ TeV and $M_{\nu_2}=1$ TeV.

\noi{\bf Figure 4.} The total cross section of the reaction
$e^-e^-\to W_2^-W_2^-Z_1$ as functions of the CM-energy
$E=\sqrt{s}$ for the masses
$M_{W_2}=0.5$ TeV,   $M_{\nu_2}=1$ TeV and for two values of the
triplet  Higgs mass, $M_{\delta}=0.8,\; 10$ TeV for the different
neutral current mixing matrices (see the text).

\noi{\bf Figure 5.} The total cross section of the reaction
$e^-e^-\to W_2^-W_2^-\g$ as functions of the CM-energy
$E=\sqrt{s}$ for the masses
$M_{W_2}=0.5$ TeV, $M_{\delta}=0.8$ TeV and $M_{\nu_2}=1$ TeV
(solid line). The cuts $\cos\theta_{\g}\leq 0.8, \; E_{\g}/E\geq
0.01$ have been applied for the scattering angle and the energy
of the photon.  Shown are separately also the contributions from
the neutrino exchange and  the Higgs exchange contributions and
their interference.

\noi{\bf Figure 6.} The total cross section of the reaction
$e^-e^-\to W_2^-W_2^-\g$ as functions of the CM-energy
$E=\sqrt{s}$ for the masses
$M_{W_2}=0.5$ TeV, $M_{\delta}=0.8$ TeV and $M_{\nu_2}=1$ TeV and
for three set of the photon cuts: $\cos\theta_{\g}\leq 0.9, \;
E_{\g}/E\geq 0.01$, $\cos\theta_{\g}\leq 0.8, \; E_{\g}/E\geq
0.01$, $\cos\theta_{\g}\leq 0.8, \; E_{\g}/E\geq 0.05$ have been
applied for the scattering angle and the energy of the photon.

\noi{\bf Figure 7.} The total cross section of the reaction
$e^-e^-\to W_2^-W_2^-\g$ as functions of the CM-energy
$E=\sqrt{s}$ for the masses
$M_{W_2}=0.7$ TeV, $M_{\delta}=0.8$ TeV and $M_{\nu_2}=1$ TeV and
for two sets of the photon cuts: $\cos\theta_{\g}\leq 0.9, \;
E_{\g}/E\geq 0.01$,  $\cos\theta_{\g}\leq 0.8, \; E_{\g}/E\geq
0.05$.

\noi{\bf Figure 8.} The total cross section of the reaction
$e^-e^-\to W_2^-W_2^-\g$ as functions of the CM-energy
$E=\sqrt{s}$ for the masses
$M_{W_2}=0.5$ TeV, $M_{\delta}=10$ TeV and $M_{\nu_2}=1$ TeV and
for
 the photon cuts $\cos\theta_{\g}\leq 0.9, \; E_{\g}/E\geq 0.01$.
Shown are separately also the contributions  from the neutrino
exchange and  the Higgs exchange contributions and their
interference.

\noi{\bf Figure 9.} The total cross section of the reaction
$e^-e^-\to W_2^-W_2^-\g$ as functions of the CM-energy
$E=\sqrt{s}$ for the masses
$M_{W_2}=0.5$ TeV, $M_{\delta}=2.0$ TeV and $M_{\nu_2}=1$ TeV and
for two set of the photon cuts: $\cos\theta_{\g}\leq 0.9, \;
E_{\g}/E\geq 0.01$, $\cos\theta_{\g}\leq 0.8, \; E_{\g}/E\geq
0.05$.

\end{document}